\documentclass[pdflatex,sn-mathphys-num]{sn-jnl}

\usepackage{graphicx}%
\usepackage{multirow}%
\usepackage{amsmath,amssymb,amsfonts}%
\usepackage{amsthm}%
\usepackage{mathrsfs}%
\usepackage[title]{appendix}%
\usepackage{xcolor}%
\usepackage{textcomp}%
\usepackage{manyfoot}%
\usepackage{booktabs}%
\usepackage{algorithm}%
\usepackage{algorithmicx}%
\usepackage{algpseudocode}%
\usepackage{listings}%
\usepackage{amsmath}
\usepackage{amsfonts}
\usepackage{bbold}
\usepackage{xcolor}
\usepackage{lmodern}

\begin{document}

\newcommand{\az}[1]{\textcolor{black}{#1}}
\newcommand{\azz}[1]{\textcolor{black}{#1}}

\title[Reinforcement learning of a biflagellate model microswimmer]{Reinforcement learning of a biflagellate model microswimmer}


\author[1]{\fnm{Sridhar} \sur{Bulusu}}\email{sridharbulusuvie@gmail.com}

\author*[1]{\fnm{Andreas} \sur{Z\"ottl}}\email{andreas.zoettl@univie.ac.at}


\affil[1]{\orgdiv{Faculty of Physics}, \orgname{University of Vienna}, \orgaddress{\street{Kolingasse 14-16}, \city{Wien}, \postcode{1090}, \country{Austria}}}




\abstract{Many microswimmers are able to swim through viscous fluids by employing periodic non-reciprocal deformations of their appendages.
Here we use a simple microswimmer model inspired by swimming biflagellates which consists of a spherical cell body and two small spherical beads representing the motion of the two flagella.
Using reinforcement learning we identify for different microswimmer morphologies quasi-optimized swimming strokes.
For all studied cases the identified strokes result in symmetric and quasi-synchronized beating of the two flagella beads.
Interestingly, the stroke-averaged flow fields are of pusher type, and the observed swimming gaits outperform previously used biflagellate microswimmer models relying on predefined circular flagella bead motion.}

\keywords{Microswimmers, Reinforcement Learning, Low-Reynolds number locomotion}



\maketitle

%
%
%
%
\section{Introduction}
\label{intro}
In nature many microorganisms are able to swim in viscous fluids at low Reynolds numbers  \cite{Lauga2009a,Elgeti2015,Zottl2016} by deforming their shape periodically non-reciprocal in time \cite{Purcell1977}.
A common self-propulsion mechanism for unicellular eukaryotic microswimmers relies on periodically deforming appendages called flagella or cilia attached at the cell body \cite{Brennen1977,Lighthill1976FlagellarHydrodynamics}.
A prominent and well-studied example is the breast-stroke-like motion of the biflagellated algae \textit{Chlamydomonas} which self-propels by the synchronous beating of two flagella in front of the cell body \cite{Goldstein2015,Gilpin2020}.
The swimming trajectory, swimming speed, power consumption and other physical properties of swimming algae are governed by low Reynolds number hydrodynamics and linked to the specific  self-generated flow field created by the active shape deformations \cite{Lauga2009a,Zottl2016,Zottl2023}.
In the absence of external forces the far-field of the fluid flow is that of a force-dipole, which in the case of the algae \textit{Chlamydomonas} have been demonstrated to oscillate in time between an extensile and a contractile force dipole field. When averaged over time the flow field is slightly contractile \cite{Guasto2010,Drescher2010,Klindt2015}, and the microswimmer is classified as a \textit{puller} (in contrast to \textit{pushers} which create an extensile flow field on average \cite{Zottl2016}).
The periodic motion of the flagella beads with respect to the cell body leads to alternating power and a recovery strokes leading to a typical "one-step-back-two-step-forward" motion where the cell body moves  backwards during the recovery stroke but significantly more forwards during the power stroke
\az{at low Reynolds number of about $\text{Re}\sim10^{-3}$.}

In the last decades various theoretical and numerical methods have been used successfully to solve the hydrodynamics around microswimmers \cite{Lauga2009a,Elgeti2015}.
\az{This has lead to a better understanding of the influence of hydrodynamic flows in, for example, the hydrodynamic trapping of swimming bacteria close to surfaces (see, e.g.\ \cite{Spagnolie2012,Schaar2015}), of relevance for the initial step of bacterial biofilm formation.
Another example is the collective dynamics of swimming microorganisms \cite{Hernandez-Ortiz2005,Saintillan2007,Zottl2014},
showing motility-induced phase separation \cite{Cates2015} or bacterial turbulence \cite{Aranson2022}.
Particularly the}
 swimming behavior of biflagellate microswimmers such as \textit{Chlamydomonas} have been investigated
to identify forces and optimum stroke patterns of the flagella \cite{Tam2011b,Bayly2011} and to understand the mechanisms for flagella synchronization.
The simplest hydrodynamic models for swimming biflagellates consist of three hydrodynamically interacting spheres, where one large sphere represents the cell body, and two smaller beads, which oscillate periodically in space in front of the body sphere, represent the flagella \cite{Friedrich2012,Polotzek2013,Bennett2013,Bennett2013b,Jibuti2017}.
Describing flagella by moving spheres have been demonstrated previously to be useful to study cilia synchronization, see e.g.\   \cite{Kotar2010,Osterman2011,Wollin2011}.
The 3-bead models for Chlamydomonas belong to the class of 3-bead triangular microswimmers consisting of three beads connected by arms or springs.
Such 3-bead triangular microswimmer models demonstrated that the leading order dipole far-field is relatively weak, and can be both of pusher and puller type. The near-field, however, for symmetric beating is governed by a rotlet doublet \cite{Rizvi2018a,Rizvi2018c,Ziegler2019}.

To change the shape of the swimmer using a 3-bead model, in principle, two approaches can be used: a shape-based approach where the shape changes are induced, and the necessary forces to maintain a certain shape change can be calculated \cite{Friedrich2012,Polotzek2013,Bennett2013}; or a force-based approach where the input of the forces are given, and the shape changes can be calculated \cite{Jibuti2017,Huebl2021b}.
Previously both approaches have been applied to 3-bead \az{Najafi-Golestanian (NG)} swimmers \cite{Najafi2005} as well as to \az{the aforementioned} triangular swimmers by applying periodic policies on either shape or forces.



\az{Since the pioneering works by, for example, Lighthill \cite{Lighthill1976} and Pironneau and Katz \cite{Pironneau1974}  about 50 years ago,
there exists great interest in understanding to which extent shape deformations of real microswimmers are hydrodynamically optimal, i.e.\ if their swimming speed or swimming efficiency can be maximized by certain shape deformations under certain constraints.
For simple model systems typically classical optimization techniques can be applied (see e.g.\  \cite{Alouges2008,Ramasamy2019,Moreau2023a}).
For example for discrete bead swimmers}
 optimization procedures have been employed e.g.\ to maximize the swimming efficiency by identifying nontrivial periodic stroke cycles \cite{Alouges2009,Wang2019c}.
Recently Reinforcement Learning (RL) \az{has} been used to identify useful strategies for shape deformations of model microswimmers.
\az{In contrast to classical methods, RL provides a powerful and often more efficient approach:
It typically relies on the usage of artificial neural networks (ANNs) which approximate a state-dependent policy for the swimmer to perform some action, i.e.\ to deform in a specific way to optimize e.g.\ the swimming speed
without the need of solving  mathematically challenging optimization procedures.
}
\az{Different} Reinforcement Learning methods have been applied to the classical three-bead \az{NG} swimmer \cite{Tsang2020,Hartl2021,Hartl2025},
generalized $N$-bead NG swimmers \cite{Tsang2020,Jebellat2024,Hartl2025}, translation and rotation of triangular swimmers \cite{Liu2021c,Huebl2021b}
3-bead hinge-swimmers \cite{Zou2022}, and generalized Purcell multi-link swimmers \cite{Qin2023c,Xiong2024},
with specific application to chemotaxis \cite{Hartl2021,Xiong2024}, context detection \cite{Zou2024}, and cooperative motion \cite{Liu2023}. 

In this work, we apply RL to a biflagellate microswimmer model where the body is connected to two effective flagella beads actuated by pairwise flagellum-body and flagellum-flagellum forces. These forces are optimized to maximize the swimming speed, under the constraint that the applied arm forces are limited to a maximum value. Our approach identifies stroke patterns, swimming trajectories and flow fields similar as experimentally observed for swimming biflagellate algae cells. Interestingly, while the time-dependent flow fields are very similar to those observed experimentally \cite{Guasto2010,Klindt2015}, the time-averaged flow field far away from the swimmer for our breast-stroke-like swimmer is that of a weak pusher instead of a weak puller, reiterating the fact that details of the stroke patterns matter if they are extensile or contractile, as seen in simple triangular microswimmer models \cite{Rizvi2018a,Rizvi2018c,Ziegler2019}.

The article is structured as follows:
In Sec.~\ref{Sec:ChlamyModel} we introduce our microswimmer model, in Sec.~\ref{Sec:RL} we present our RL approach.
The results are presented in Sec.~\ref{Sec:Results}, followed by a conclusion in Sec.~\ref{Sec:Con}.

\section{Microswimmer model}
\label{Sec:ChlamyModel}

Our biflagellate microswimmer is inspired by the unicellular algae \textit{Chlamydomonas} 
which is a common model organism for studying flagellar motion, synchronization and cell motility \cite{Goldstein2015}.
Chlamydomonas algae have two periodically beating flagella in front of their cell body which allow the cell to self-propel forward.

\begin{figure}
\centering
\includegraphics[width=0.5\textwidth]{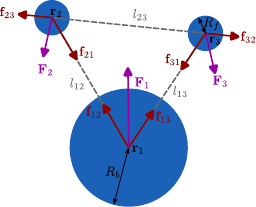} 
\caption{Sketch of the 3-bead biflagellate microswimmer model: The cell body is represented by a large \az{bead} of radius $R_b$, and the two flagella by two smaller beads of radius $R_f$. The positions of the beads, $\mathbf{r}_i$ change by applying forces $\mathbf{F}_i$ onto them, realized as a combination of pairwise arm forces $\mathbf{f}_{ij}$ \az{between arms of length $l_{ij}$,} which automatically ensure force- and torque-free conditions.}
\label{fig:1}       
\end{figure}

In our model the main bead is considered to be the cell body  with  radius $R_b$ located at position $\mathbf{r}_1(t)$, and the other two
smaller beads of radius $R_f<R_b$ located at $\mathbf{r}_2(t)$ and $\mathbf{r}_3(t)$ are used to model the flagella (see Fig.~\ref{fig:1}).
The bead positions change over time $t$ by time-dependent forces $\mathbf{F}_i(t)$, $i=1,2,3$, acting on the respective beads \cite{Rizvi2018a,Ziegler2019,Hartl2021,Huebl2021b,Hartl2025}.
This leads to time-dependent distances between the beads $l_{ij}(t) \equiv l_{ji}(t)  = |\mathbf{r}_i(t) - \mathbf{r}_j(t)|$, $i < j$.
Hence, in the frame of reference of the swimmer its shape can be characterized solely by the three bead distances $l_{12}$, $l_{13}$ and $l_{23}$.
Forces $\mathbf{F}_i(t)$ are not applied directly onto the beads but are modeled as virtual \textit{muscle arm forces} $\mathbf{f}_{ij}(t)$ between all three pairs of beads,
acting along the distance vectors of the beads, i.e.\ $\mathbf{f}_{ij}(t) = f_{ij}(t) \hat{\mathbf{r}}_{ij}(t)$, $i \neq j$, with
$\hat{\mathbf{r}}_{ij}(t) = (\mathbf{r}_i(t) - \mathbf{r}_j(t))/l_{ij}(t)$ and $f_{ji}\equiv f_{ij}$.
Hence, the forces on the beads are controlled solely by three scalar arm force strengths, $f_{12}$, $f_{13}$ and $f_{23}$.
Since $\mathbf{f}_{ji}(t) = - \mathbf{f}_{ij}(t)$, this automatically ensures that the swimmer is moving force- and torque-free \cite{Rizvi2018a}.
The total force on each bead is then the sum of the two connecting arm forces, $\mathbf{F}_i(t) = \sum_{j \neq i} \mathbf{f}_{ij}(t) $ with
$\sum_{i=1}^3 \mathbf{F}_i(t) = 0$ and $\sum_{i=1}^3 \mathbf{r}_i(t) \times \mathbf{F}_i(t) = 0$.

Each of the three arm forces $f_{ij}(t)=f_{ij}^a(t)+f_{ij}^r(t)$, $j<i$, consist of two contributions: First, the arms are extended or compressed actively by muscle forces $f_{ij}^a(t)$.
These forces are not predefined but are proposed by an artificial neural network \textit{via} reinforcement learning, as described in Sec.~\ref{Sec:RL}.
They are limited in magnitude, by a maximum force amplitude $f_{\text{max}}$ such that $-f_{\text{max}} \le f_{ij}^a(t) \le f_{\text{max}}$.
\az{The specific value of $f_{\text{max}}$ sets the force-scale in the system, an therefore
also the velocity scale of the microswimmer because of the linear relation  between velocities and forces at low Reynolds number.
This} in turn sets the time scale and hence the stroke frequencies scale as $\az{\propto} f_{\text{max}}$.
As a consequence it automatically sets a limit on the maximum power the microswimmer can use to swim,
\az{which scales as $\propto f_{\text{max}}^2$.
Hence, the specific value of $f_{\text{max}}$ does not influence the optimum stroke kinematics but only the scaling of force, velocity, time etc.}

Second, passive restoring forces $f_{ij}^r(t)$ are applied when the bead distances become too large or too small:
We assume that the beads are connected by massless arms which are allowed to extend \az{freely} 
only up to a certain length $l_{bf}^\text{max}$ between the body bead and the flagella beads, and up to a length $l_{ff}^\text{max}$ between the two flagella beads.
Similarly, arm compression is constrained by the respective minimum arm distances $l_{bf}^\text{min}$ and $l_{ff}^\text{min}$.
Extending or contracting  above or below these values \az{is restricted} 
 by harmonic forces of the form (see also \cite{Hartl2021,Hartl2025,Huebl2021b})
\begin{equation}
    f_{1j}^r(t) = \begin{cases}
k(l_{1j}(t)-l_{bf}^\text{min}) & \text{for } l_{1j}(t)<l_{bf}^\text{min} \\
k(l_{1j}(t)-l_{bf}^\text{max}) & \text{for } l_{1j}(t)>l_{bf}^\text{max} \\
0 & \text{else}
\end{cases}
\end{equation}
for $j=2,3$, i.e.\ between body and flagella beads, and
\begin{equation}
    f_{23}^r(t) = \begin{cases}
k(l_{23}(t)-l_{ff}^\text{min}) & \text{for } l_{23}(t)<l_{ff}^\text{min} \\
k(l_{23}(t)-l_{ff}^\text{max}) & \text{for } l_{23}(t)>l_{ff}^\text{max} \\
0 & \text{else}
\end{cases} 
\end{equation}
between the two flagella beads, where $k$ is a spring constant.
\az{
These forces are thus expanding the arms if they become smaller than the desired minimum values $l_{bf}^\text{min}$ and $l_{ff}^\text{min}$,
and contracting if they become larger than the desired maximum values $l_{bf}^\text{max}$ and $l_{ff}^\text{max}$.
The spring constant is chosen sufficiently stiff such that arm expansion and contraction is almost limited to the minmum and maximum values given above.}

In our approach we neglect the rotation of the beads and only take bead translation into account.
Furthermore we neglect thermal fluctuations.
Then at low Reynolds number the relation between the bead velocities $\mathbf{v}_i$ and the forces $\mathbf{F}_i$ on the beads are linear,
\begin{equation}
   \mathbf{v}_i = \sum_{j=1}^3 \mathbb{M}_{ij} \cdot \mathbf{F}_j
   \label{eq:v}
\end{equation}
where $\mathbb{M}_{ij}$ is the mobility tensor which is symmetric $\mathbb{M}_{ij} = \mathbb{M}_{ji}$ \cite{KimKarila}.
The self mobilities  $\mathbb{M}_{ii}$ are given by $\mathbb{M}_{11} = 1/(6\pi\eta R_b)\mathbb{1}$ and $\mathbb{M}_{22} = \mathbb{M}_{33} = 1/(6\pi\eta R_f)\mathbb{1}$
where $\eta$ is the fluid viscosity.
The cross-mobilities $\mathbb{M}_{ij}$, $i \ne j$ capture the hydrodynamic interactions between the beads, which are required for the microswimmer to obtain any net movement in the absence of external forces.
We consider here the cross mobilities in the Rotne-Prager approximation,
\az{applicable to describe far-field hydrodynamic interactions in the zero Reynolds number limit}  \cite{KimKarila},
\begin{equation}
    \mathbb{M}_{ij}  =  \frac{1}{8\pi\eta |\mathbf{r}_{ij}|}\left[ \mathbb{1}+\frac{\mathbf{r}_{ij} \otimes \mathbf{r}_{ij}}{|\mathbf{r}_{ij}|^2}  \\
      + \frac{R_i^2+R_j^2}{|\mathbf{r}_{ij}|^2}\left( \frac{1}{3}\mathbb{1} - \frac{\mathbf{r}_{ij} \otimes \mathbf{r}_{ij}}{|\mathbf{r}_{ij}|^2} \right)   \right]
      \label{eq:M}
\end{equation}
with $\mathbf{r}_{ij} = \mathbf{r}_i - \mathbf{r}_j$ and $R_i$ is the radius of bead $i$.
While we consider three-dimensional hydrodynamics, we allow initial bead positions and  forces applied only in the $x$-$y$ plane, such that the motion of the swimmer is two-dimensional.

In the simulations we use the following parameters:
The viscosity is set to $\eta=1$, the radius of the flagella bead is $R_f=1$ and we use different sizes of the cell body, $R_b/R_f= \{3,4,5,6,7,8 \}$.
The maximum arm force is set to $\az{f_\text{max}}=10$, and the spring constant of the restoring forces is $k=10$.
This allows the arms to extend approximately to lengths $l_{bf}^\text{max}+1$ and $l_{ff}^\text{max}+1$, and to contract approximately to lengths
$l_{bf}^\text{min}-1$ and $l_{ff}^\text{min}-1$.
The minimum distances are chosen such that the beads do not \az{collide or} overlap, \az{which is fulfilled by using}  $l_{bf}^\text{min} = R_b + 3R_f$ and $l_{ff}^\text{min} = 4R_f$.
The maximum distances are related to the length $L_f$ of the effective flagella, which we chose to be $L_f=3R_b$, motivated by the typical physiology of Chlamydomonas \cite{Goldstein2015}.
Thus we use the maximum body-arm extension $l_{bf}^\text{max} = R_b-R_f+L_f/2$ where body-flagella distance is approximately limited below half of the flagella length.
For the maximum distances between the flagella beads we set $l_{ff}^\text{max} = L_f$\az{, thus allowing the flagella beads to be separated by a distance comparable to real maximum flagella-flagella distances observed in swimming biflagellates.}
The equations of motion for the bead positions $\frac{\mathrm{d}\mathbf{r}_i(t)}{\mathrm{d}t} = \mathbf{v}_i(t)$ are solved by simple Euler integration using Eq.~(\ref{eq:v}) with time step $\Delta t = 0.026 \tau$ where $\tau = 2R_f/v_0=12\pi\eta R_f^2/ \az{f_\text{max}}$ is the time to move a flagella bead through the fluid with maximum force $\az{f_\text{max}}$ for a bead diameter distance,
at velocity $v_0=f_\text{max}/(6\pi\eta R_f)$.

\section{Reinforcement Learning}
\label{Sec:RL}

In general Reinforcement Learning (RL) is a method in the area of Machine Learning and Control which finds policies of agents which act and interact with an environment in order to achieve a particular task \cite{Shakya2023}. The policy is the way the agent acts depending on environmental variables, and it is quantified as a function from the input of the environment  to the control outputs. This function is for example approximated by an  Artificial Neural Network (ANN) (see also Appendix~\ref{A:ANN}) identified by the RL algorithm. 
The success of the task is quantified with the help of a reward signal: If the task is successfully accomplished then a positive or high reward is associated with the corresponding policy. By trial and error the ANN is modified in order to optimize the performance, i.e. to find a good reward signal. 
The ANN is adapted in a sequence of generations and the learning is successful when the reward signal converges to its optimum.



We make use of a RL algorithm called Neuroevolution of Augmenting Topologies (NEAT) \cite{Stanley2002a} which optimizes not only the weights and biases of the ANN but also its topology.
Neuroevolution (NE) is a class of  methods which use evolutionary algorithms (EAs) for generating \az{and modifying not just a single ANN but a population of} ANNs during training. In conventional methods the topology and structure of artificial neural networks are fixed and methods like gradient descent are used to optimize the weights and biases. The specialty of NEAT is that the topology of the artificial neural network is modified over the course of many generations. 
This makes the method robust in finding quasi-optimum solutions.
\az{We refer here and in the following in this paper to \textit{quasi-optimum} rather than to \textit{optimum} solutions.
The reason is that because of the optimization procedure with ANNs, and the infinite action space, strictly speaking no unique optimum solution of the ANN parameters can be found, but rather a potentially infinite set of different ANNs which after a certain amount of generations converge to the same quasi-optimum solution, but which have different ANN topology and weights.
}
More details about the NEAT algorithm can be found in Appendix~\ref{A:NEAT} and in Ref.~\cite{Stanley2002a}.

In our case the agent is the microswimmer consisting of three beads and its ANN, and the inputs are its arm lengths $l_{12}$, $l_{13}$ and $l_{23}$. 
Therefore, the environment is characterized by the internal shape of the swimmer, propagating in a viscous fluid,
and the output is represented by the arm forces $f_{12}$, $f_{13}$ and $f_{23}$.
\az{O}ur aim is to find quasi-optimum policies for the three arm force\az{s, depending only on the} 
internal state of the swimmer, \az{i.e.}\ the arm lengths $l_{12}$, $l_{13}$ and $l_{23}$. 
ANNs are then used 
to identify nonlinear functions for the forces $f_{ij}(\{ l_{ij} \})$ to maximize the \az{reward, as outlined below}.

In the following we consider two cases, which we refer to the \textit{symmetric} (S) and \textit{non-symmetric} (N) RL scheme: 
In the symmetric (S) case we assume that the microswimmer performs perfectly symmetric motion of the flagella beads with respect to the cell body, where the flagella beads move in synchrony. This is the case when the two body-flagella arm lengths are the same at all times, $l_{12}(t) = l_{13}(t)$\az{,} realized by restricting the body-flagella arm forces to be the same, $f_{12}(t) = f_{13}(t)$. Hence, for the symmetric (S) case the output of the ANN consists of only two values, one for  $f_{12} \equiv f_{13}$ and one for $f_{23}$.
In contrast, for the non-symmetric (N) case we allow the microswimmer to modify its shape without the aforementioned constraint, where it is allowed to learn strategies which do not rely on synchronized flagella motion.





Training an artificial neural network takes place over a specified number of generations.
The NEAT algorithm accommodates for a hyperparameter called population size which keeps the number of competing \az{ANNs} to a certain size, in our case to 100\az{, comparable to the population size used in our previous work \cite{Hartl2021}}.
\az{The population size has been chosen sufficiently large to allow the development of distinct sub-populations to foster diversity \cite{Stanley2002a}, but sufficiently small to reduce computational cost, since the simulation time scales approximately linearly with the population size.}
In each generation this population of ANNs proposes 100 different policies and for each individual genome of the population the reward (performance) is calculated.
After each generation through selection and mutation a new population of ANNs is constructed and again their performance evaluated.
In our training we perform training over $2000$ generations to ensure that a quasi-optimum solution has been found.

For each generation and each individual swimmer of the population the equations of motion 
are  integrated for $N_t=10^4$ time steps using Eq.~(\ref{eq:v}).
\az{In each time step the forces on the arms are calculated based on the output of the active forces $f_{ij}^a$ from the ANN, which depend \textit{via} a non-linear function on the instantaneous arm lengths $l_{ij}$ at this time.
To be concrete, the specific ANNs for the swimmer population, and therefore the specific mapping of the instantaneous arm lengths $l_{ij}$ to the arm forces $f_{ij}$, are fixed during the entire training run at a given generation, but are modified in subsequent generations.
}

The initial position of the body is chosen to be at the origin, $\mathbf{r}_1^0 = (0,0)$, and the initial positions $\mathbf{r}_2^0$ and $\mathbf{r}_3^0$ of the  flagella beads are given by $x_2^0 = x_3^0 = \sqrt{(l_{12}^0)^2 - (l_{ff}^\text{min}/2)^2}$ and $y_2^0 = l_{ff}^\text{min}/2$, $y_3^0 = \az{-}l_{ff}^\text{min}/2$.
where $l_{12}^0=(l_{bf}^\text{min}+l_{bf}^\text{max})/2$.
A small random noise is added onto each coordinate of each bead to obtain robust training results.
During training, the
\az{\textit{reward}} 
 of an individual swimmer with a certain neural network is
 calculated as the traveled distance of the cell body
in $x$ direction over a time period $T_T=N_t \Delta t$, i.e.\ by 
\az{$x_1(T_T) - x_1^0$.}
\az{We then define the \textit{fitness} of the swimmer as the traveled distance per simulation time $T_T$, i.e.\ as
$(x_1(T_T) - x_1^0)/T_T$, which is simply the mean velocity of the swimmer obtained during training, and thus indeed concides with the goal of our optimization procedure: The NEAT algorithm will identify the fittest (i.e.\ the fastest) swimmer for each generation, and further improve the fitness to find a quasi-optimized solution after 2000 generations.
}



\begin{figure}
\centering
  \includegraphics[width=\textwidth]{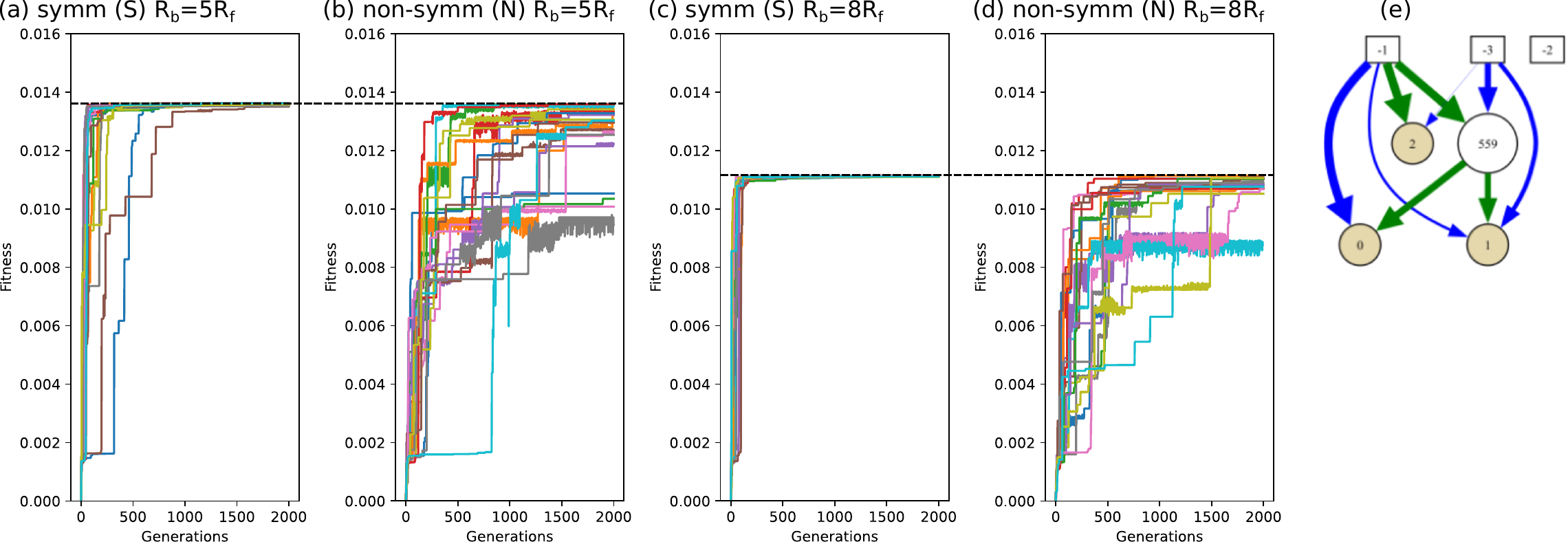} 
\caption{(a-d) shows each 20 independent neuroevolution runs of the best fitness from a population of 100 ANNs depending on the number of generations of RL cycles. 
The dashed black lines indicate the overall best fitness from the 20 neuroevolution runs of the symmetric (S) swimmers after 2000 generations. (e) Example of a neural network, shown here for the fittest solution for $R_b=6R_f$ using the intrinsic plotting function of NEAT \cite{Stanley2002a}. Inputs: [-1]: $l_{12}$, [-2]:$l_{13}$, [-3]:$l_{23}$; outputs (yellow circles): \az{(0)}: $\az{f_{12}^a}$, \az{(1)}: $\az{f_{13}^a}$, \az{(2)}: $\az{f_{23}^a}$.
\az{Blue arrows correspond  to negative weights, and green arrows to positive weights. The arrow thickness visualizes the absolute value of the weights.}
}
\label{fig:B}       
\end{figure}

Because of the inherent stochastic nature of the NEAT algorithm we perform the neuroevolution process 20 times for each set of parameters  to investigate the convergence to quasi-optimum solutions. 
Figure~\ref{fig:B}(a-d) shows the evolution of the  best fitness of the population depending on the number of generations
for two different body sizes $R_b/R_f$ both for the symmetric (S) and non-symmetric (N) case.
 We can see that for the symmetric (S) cases  the  fitness converges more robustly towards the \az{quasi-}optimum,
 whereas for the non-symmetric (N) swimmer some of the 20 neuroevolution processes do not converge; however, in all cases at least several of them converge and a quasi-optimum solution can be found, which, interestingly, is very close to those obtained for the symmetric (S) swimmers (see also black dashed lines in Fig.~\ref{fig:B}(a-d)) \az{but always slightly smaller.}
 We thus conclude that learning the optimum policy is easier for the symmetric (S)  case, which is not surprising considering the fact that the swimmer body motion is by construction only in 1D and optimization occurs faster.
 We also note that there is  noise on top of the fitness curves, because the fitness is reevaluated after each generation with randomized initial conditions of the bead positions. 
  For evaluation  the training results of the best neuroevolution run are adopted for further analysis presented below in  Sec.~\ref{Sec:Results}.

A typical neural network  obtained by NEAT is shown in Fig.~\ref{fig:B}(e) for the non-symmetric case for $R_b=6R_f$. In this case the network only consists of a single hidden neuron. The arm length $l_{13}$ is not used at all by the network. The policies represented by the ANNs will be  discussed later in Sec.~\ref{sec:policy}.

\begin{figure}
\centering
  \includegraphics[width=0.5\textwidth]{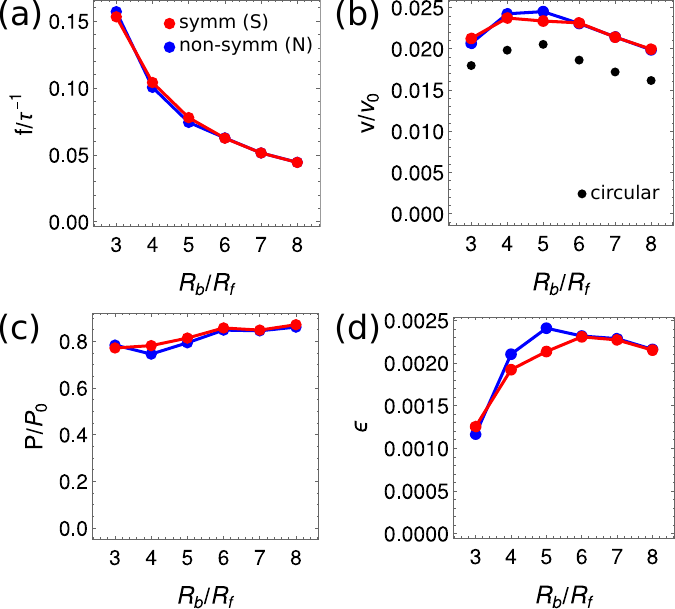} 
\caption{(a) Stroke frequency, (b) stroke-averaged swimming speed, (c) stroke-averaged power consumption, and (d) hydrodynamic swimming efficiency for non-symmetric (N) and symmetric (S) swimmers depending on the cell body size $R_b$.}
\label{fig:2}       
\end{figure}

\section{Results}
\label{Sec:Results}

\subsection{Swimming performance}
\label{sec:perform}
After training of the different microswimmers we analyzed the performance of the swimmers using the selected best policies per swimmer type by monitoring the bead positions and forces over $3000$ time steps, using the same conditions as in the RL training runs.
For all our microswimmers we observe that the motion of the beads, and henceforth of the arm lengths, are periodic in time, after a short initial transient.
Interestingly, we observe that the non-symmetric (N) swimmers adapt a quasi-symmetric motion of the flagella beads, which is very similar to the motion of the respective symmetric (S) microswimmers. 
Therefore the results for symmetric (S) and non-\az{symm}etric (N) swimmers are very similar, as we have already seen for the fitness in Fig.~\ref{fig:B}(a-d).
We first determined the frequencies $f$ of the periodic motions,
which decreases with increasing body size $R_b$ similarly for (N) and (S) microswimmers, see Fig.~\ref{fig:2}(a).

Because the solutions are (almost) symmetric also for the non-symmetric (N) swimmers, the body bead moves (almost) only in the $x$ direction.
We then determine the swimming velocities in the $x$ direction of the microswimmers as the stroke-average distance traveled by the cell body,
$v = (x_1(t_0 + T) - x_1(t_0))/T$ with $T=1/f$ the stroke period, which holds $\forall t_0$ for periodic motion.
Note that the stroke-averaged velocit\az{ies} deviate to some extent from the fitness of the swimmers in the training runs due to the lack of knowledge of $T$ in the latter. 
\az{They are plotted in Fig.~\ref{fig:2}(b) for different body sizes $R_b$, and show}
a local maximum at a certain range of body sizes.
Again the swimming speed for a given body size $R_b$ is comparable for the symmetric (S) and non-symmetric (N) case.
\az{Interestingly, although the fitness obtained during training are always (slightly) higher for the symmetric (S) swimmers, for $R_b=4R_f$ and $R_b=5R_f$
the stroke-averaged swimming velocities $v$ are slightly larger for the non-symmetric (N) swimmer. This discrepancy stems from the fact that the stroke period $T$ was not known a-priori, and the training time $T_T$ was thus not a multiple integer value of $T$, reiterating the fact that optimization of mean speed for a finite time $T_T$ is not exactly the same as optimizing the stroke-averaged velocity $v$.}
\az{These two measures would coincide for $T_T \rightarrow \infty$. Practically, since in our approach $T_T \gg T $, i.e.\ at least $T_T \gtrsim 11.8T$ for the largest stroke periods, the mismatch between them is expected to be  small.}

\az{We can roughly compare our results to real swimming \textit{Chlamydomonas} when mapping our simulation units to physical units.
The radius of Chlamydomonas cells is typically $\sim 5\mathrm{\mu} \text{m}$, which we map to our typical model bead size $R_b=5R_f$. 
 Flagella forces are on the order of $\sim 10\,\text{pN}$ or even smaller \cite{Bayly2011}, which we map to $f_\text{max}$,
 and water viscosity is $\sim 10^{-3}\,\text{Pa\,s}$.
 Using these mappings we obtain $\tau \rightarrow 0.0037s$ and $v_0 \rightarrow 530\mu \text{m\,s}^{-1}$.
 With the data shown in Fig.~\ref{fig:2}(a,b) for $R_b = 5R_f$ we get from our rough estimate a corresponding stroke frequency $f$ of $\sim 21 \text{s}^{-1}$, which is indeed comparable but smaller than the actual typical beating frequency of Chlamydomonas of $\sim 50 \text{s}^{-1}$.
 The obtained swimming speed corresponds to $\sim 11\,\mu \text{m}\, \text{s}^{-1}$. This is about one order of magnitude smaller than for real swimming Chlamydomonas. A possible reason is our simple model which approximates the flagella by single beads, and is lacking anisotropic friction and near-field effects of real slender and shape deforming flagella.
}


We further determine the stroke-averaged power consumption 
\begin{equation}
\mathcal{P}    = \frac{1}{T}\sum_{j=1}^{N_S} \sum_{i=1}^3 \mathbf{v}_i(j\Delta t) \cdot \mathbf{F}_i(j\Delta t) \Delta t  
\end{equation}
where $\mathbf{v}_i(t)$ and $\mathbf{F}_i(t)$ are the obtained instantaneous bead velocities and forces, respectively,
and $N_S=T/ \Delta t$.
These are compared with the maximum power $\mathcal{P}_0 = \frac{f_\text{max}^2}{6\pi\eta}\sum_i R_i^{-1}$ obtained for three non-interacting spheres independently dragged with force $f_\text{max}$.
As shown in Fig.~\ref{fig:2}(c) the power consumption is comparable for all systems, operating close to highest possible power. This can be seen below where we discuss the actual policies and arm forces $f_{ij}$, which are in magnitude always close to the maximum value $f_\text{max}$.
\az{Small visible differences can be identified for different values of $R_b$, stemming from different hydrodynamic bead-bead interactions which influence the actual bead velocities, as well as the presence of restoring forces.}

Finally we calculate the swimming efficiency \cite{Lauga2009a,Hartl2025} $\epsilon = (6\pi\eta v^2 \sum_{i}R_i)/ \mathcal{P}$.
Since all the swimmers operate close to quasi-maximum power, the swimming is relatively efficient up to approximately $0.3\%$ see Fig.~\ref{fig:2}(d).
\az{As can be seen in Fig.~\ref{fig:2}(d), although we did not optimize for the swimming efficiency, the shape of the curve is comparable to the swimming velocity curve shown in Fig.~\ref{fig:2}(b), however not exactly because of the small differences in power consumption, see Fig.~\ref{fig:2}(c).}
The \az{efficiency} values from our approach are lower, but on the same order of magnitude to the optimized swimming gaits of more accurate hydrodynamic models for biflagellates by Tam and Hosoi \cite{Tam2011b}, who obtained  efficiencies of up to $0.8\%$ for breast-stroke like motion.

\begin{figure}
\centering
  \includegraphics[width=0.9\textwidth]{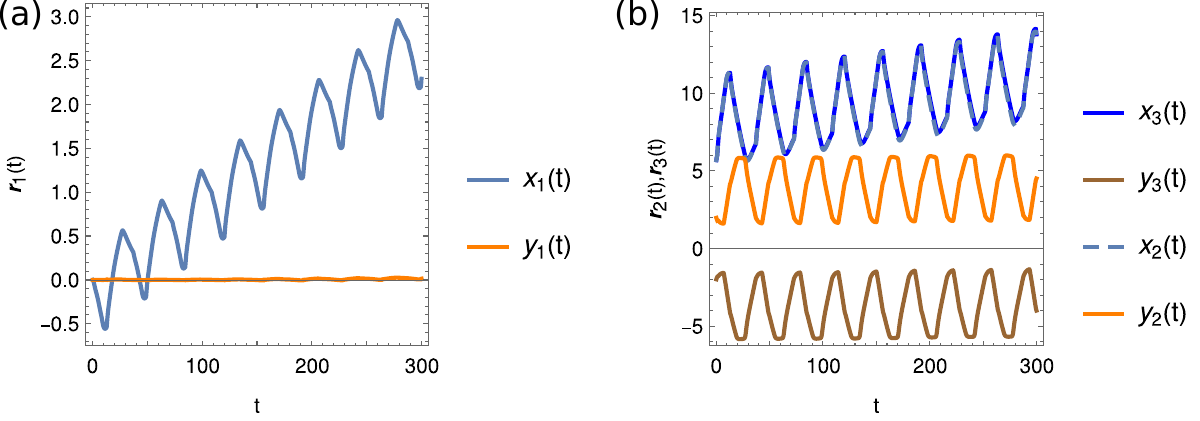} 
\caption{(a) Trajectory of the body for $R_b=5R_f$ \az{for a non-symmetric (N) swimmer}. (b) Corresponding trajectories of the two flagella beads.}
\label{fig:A}       
\end{figure}

\subsection{Swimming gaits}
\label{sec:gaits}
Next we analyze the obtained swimming gaits.
As mentioned, the swimming stroke patterns of our optimized non-symmetric (N)-microswimmers are almost symmetric, very similar to the perfectly symmetric solutions of the intrinsically symmetric (S)-microswimmers.
 This quasi-symmetric solutions thus ensure that the (N)-microswimmers also move almost only along the $x$ axis.
 A typical quasi-symmetric solution obtained for a non-symmetric (N) swimmer can be seen in Fig.~\ref{fig:A}  for  body size  $R_b=5R_f$. Here
  the body almost moves on a straight line in the $x$ direction with negligible motion in the $y$ direction (Fig.~\ref{fig:A}(a)).
 The $x$ position of the body performs the typical "\az{one}-step-back-\az{two}-step-forward" motion as known for \textit{Chlamydomonas}.
 The flagella beads move periodically in the $x$-$y$ plane.
 The fact that $x_3(t) \approx x_2(t)$ (curves almost indistinguishable e.g.\ in Fig.~\ref{fig:A}(b)) and $y_3(t) \approx -y_2(t)$ reiterates the fact that the swimmer performs almost perfectly synchronized periodic motion.

\begin{figure}
\centering
  \includegraphics[width=0.9\textwidth]{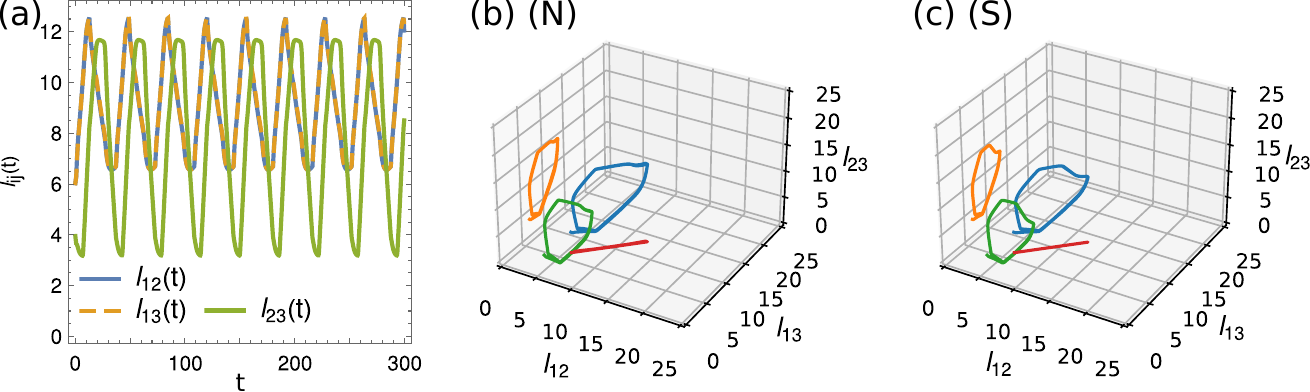} 
\caption{Periodic arm motion for the non-symmetric (N) and symmetric (\az{S}) microswimmer for $R_b=5\az{R_f}$. (a) Motion of the time-dependent arm lengths $l_{ij}(t)$ \az{shown here for the non-symmetric (N) swimmer}. (b,c) Corresponding arm length motion in the arm length phase space \az{for both (N) and (S) swimmers}. Blue: Actual trajectory: Red, orange, green: Projections to the 2D subspaces.}
\label{fig:3}       
\end{figure}

This can further be illustrated by plotting the arm lengths $l_{ij}$, as shown in Fig.~\ref{fig:3}(a), which
vary periodically, and  the distances of the body to the two flagella beads are approximately the same,
$l_{12} \approx l_{13}$.
Fig.~\ref{fig:3}(b) shows all arm lengths $l_{ij}$ in the corresponding phase space, again indicating the periodic motion in the swimmer shape space (blue curve).
The projections onto the 2D sub-spaces are shown in orange, green and red.
The deformations in shape space compare very well with the fully symmetric (S)-microswimmers, shown in Fig.~\ref{fig:3}(c).


\subsection{Flow fields and dipole strength}
\begin{figure}
\centering
  \includegraphics[width=\textwidth]{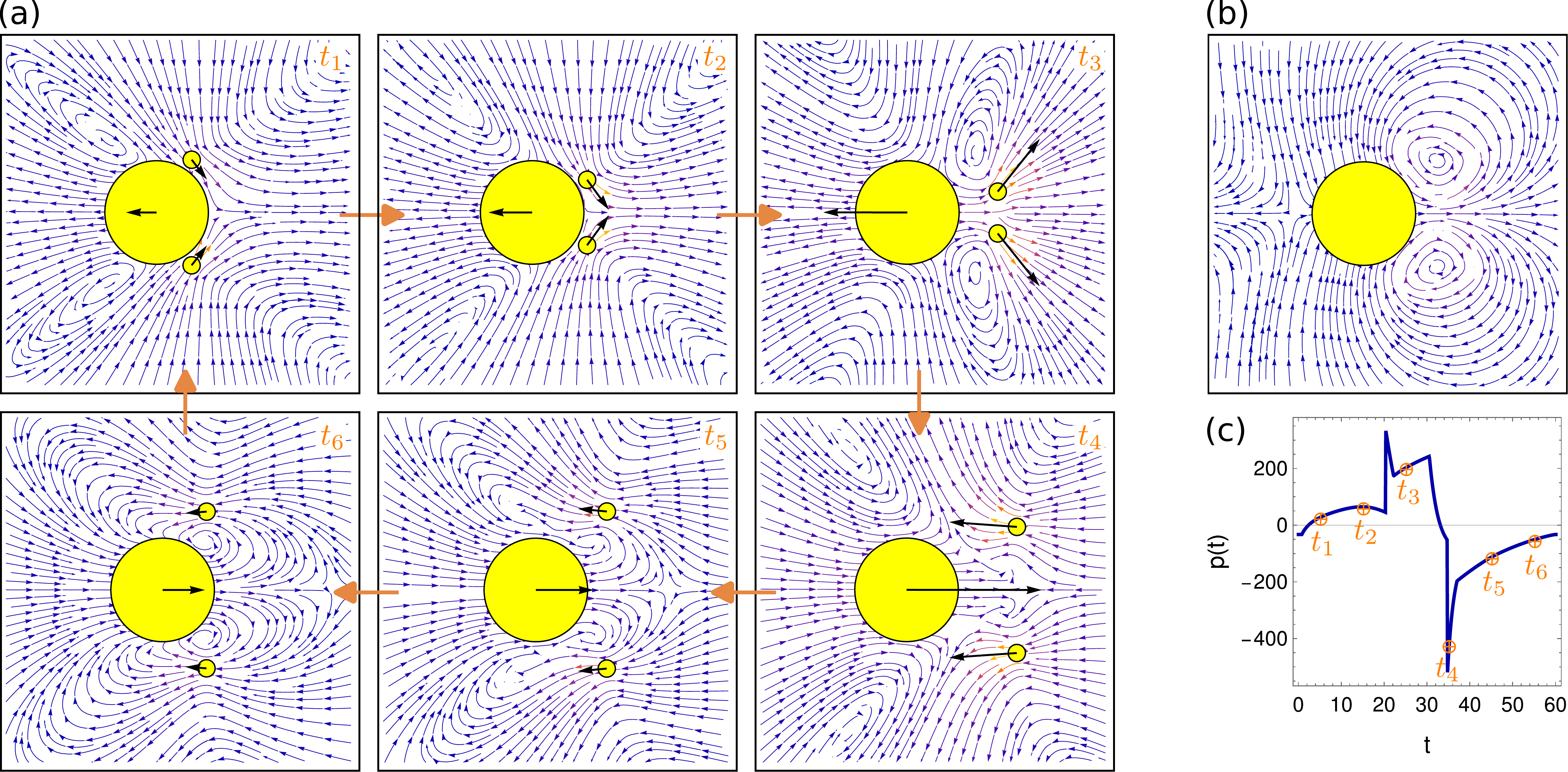} 
\caption{(a) Six snapshots \az{at times $t_i=i T/6$, $i=1,\dots,6$,} of the stroke cycle of a biflagellate microswimmer consisting of cell body of radius $R_b$ (large yellow bead) and two flagella beads of radii $R_f$ (small yellow beads), the instantaneous \az{total} forces \az{$\mathbf{F}_i$} on the beads (black arrows), and the flow field stream lines created by the microswimmer. (b) Stroke-averaged flow field. (c) Time-dependent instantaneous force dipole strength over a full stroke cycle. The six orange crosses correspond to the snapshots shown in (a).}
\label{fig:5}       
\end{figure}

 The motion of the beads creates a dynamic flow field pattern $\mathbf{v}_f(x,y,t)$, as shown in Fig.~\ref{fig:5}(a) for six instantaneous snapshots within a stroke cycle, together with the instantaneous forces $\mathbf{F}_i(t)$  on the beads for $R_b=6R_f$. 
 Shown is the flow field in the Oseen approximation, created by the three beads, given by
\begin{equation}
  \mathbf{v}_f(\mathbf{r}(t))= \sum_{i=1}^3 \mathbf{v}_f^i(\mathbf{r}(t))
\end{equation}
where
\begin{equation}
    \mathbf{v}_f^i(\mathbf{r}(t)) = \mathcal{S}_i(\mathbf{r};\mathbf{r}_i) \cdot \mathbf{F}_i
\end{equation}
is the flow created by bead $i$
with
\begin{align}
    \mathcal{S}_i (\mathbf{r};\mathbf{r}_i) = & \frac{1}{6\pi |\mathbf{r}-\mathbf{r}_i|} \left( \mathbb{1} + \frac{(\mathbf{r}-\mathbf{r}_i) \otimes (\mathbf{r}-\mathbf{r}_i)}{|\mathbf{r}-\mathbf{r}_i|^2}  \right) \\
    & + \frac{R_i^2}{6\pi |\mathbf{r}-\mathbf{r}_i|^3} \left( \frac 1 3 \mathbb{1} - \frac{(\mathbf{r}-\mathbf{r}_i) \otimes (\mathbf{r}-\mathbf{r}_i)}{|\mathbf{r}-\mathbf{r}_i|^2}  \right) \, ,
\end{align}
representing the stokeslet and source-dipole contributions to the flow field for different times $t_i=i T/6$, $i=1,\dots,6$.
\az{Similar as for real swimming Chlamydomonas, our microswimmers perform \textit{power strokes} where they move forward, alternating with \textit{recovery strokes} where they move backward, visualized in the second and first row in Fig.~\ref{fig:5}(a), respectively.
The created thrust force is larger during the power stroke, leading to an overall positive net displacement after one stroke cycle,
again, similar to that observed experimentally \cite{Guasto2010,Drescher2010,Klindt2015}:
During the power stroke, the flagellar beads are pulled towards the cell body, leading to a forward motion of the body, which in turn pulls against the flagella beads.
Here a contractile puller-fluid flow is created instantaneously, as shown at time snapshots $t_4-t_6$,
where the fluid velocity field points inwards along the swimmer axis, and outwards perpendicular to it \cite{Lauga2009a,Zottl2016}.
}
\az{In contrast, for the recovery stroke, as shown in the first row in Fig.~\ref{fig:5}(a),} the far-field of the fluid velocity is that of a pusher, i.e.\ it is extensile, where the flagella in front of the body push fluid away from the body, and the body is pushing backwards in the other direction.
\az{Indeed the connection between power stroke and puller flow field, and between recovery stroke and pusher flow field resembles the experimental observations}
of the biflagellate microswimmer \textit{Chlamydomonas reinhardti}
\cite{Guasto2010,Drescher2010,Klindt2015}.

The stroke-averaged fluid velocity flow field is shown in Fig.~\ref{fig:5}(b)\az{, which looks s}imilar
 to that reported experimentally for swimming \textit{Chlamydomonas} algae \az{: The}
time-averaged near field
is governed by a rotlet-doublet flow field, induced by the rotation and counter-rotation of the two flagella beads \az{\cite{Guasto2010,Drescher2010,Klindt2015}}.
Indeed it have been shown using simple triangular microswimmers that the dominant contribution to the flow near the microswimmer for synchronous beating is that of a rotlet-doublet \cite{Rizvi2018c,Ziegler2019}.
Only far away from the swimmer the leading order stroke-averaged flow field is a force dipole, and is in the case of \textit{Chlamydomonas} algae  a (weak) puller flow field \cite{Guasto2010,Drescher2010,Klindt2015}.

We calculated the instantaneous force dipole strengths $p(t)$ created by the microswimmers,
approximated by
$p(t) = \mathbf{F}_1(t) \cdot \mathbf{d}(t)$ where $\mathbf{d}(t) = (\mathbf{r}_1(t) - (\mathbf{r}_2(t) + \mathbf{r}_3(t))/2)$   for synchronously beating flagella beads where the flagella-body distance $\mathbf{d}(t)$ is to a very good approximation aligned along the direction of the opposing force pairs $\mathbf{F}_1(t)$ and $\mathbf{F}_2(t)+\mathbf{F}_3(t)$.
The instantaneous force dipole strengths $p(t)$ for $R_b=6R_f$ is plotted in Fig.~\ref{fig:5}(c), where the corresponding six snapshots from Fig.~\ref{fig:5}(a) are highlighted in orange. The instantaneous values of the force dipole strength are  $p(t) >0$ during the recovery stroke, as expected when moving in the pusher mode, and $p(t)<0$ during the power stroke, where the microswimmer moves in the puller mode.

Interestingly, when averaged over a stroke cycle the force dipole strength is for all our considered microswimmers slightly positive,
$\bar{p} = \frac{1}{T}\int_0^T p(t)\mathrm{d}t >0$,
\az{meaning that in the far field our microswimmers behave as pushers,}
in contrast to the \az{time-averaged} puller \az{flow field of} \textit{Chlamydomonas}.
This can further be seen in Fig.~\ref{fig:5}(b), where a stagnation point in the fluid is located behind the cell body bead, leading to an extensile flow field far away from the swimmer. In contrast, the stagnation point of the time-averaged flow field of the swimming puller algae  \textit{Chlamydomonas} is located at the front of the cell body, leading to a contractile flow field \cite{Guasto2010,Drescher2010,Klindt2015}.
\az{So all in all our flow field solutions capture well the important stroke-averaged near field of swimming  \textit{Chlamydomonas}, which is a rotlet doublet flow,
but our flow field far away from the swimmer shows it is a weak pusher instead of a weak puller.
}
\az{Our work demonstrates that breast-stroke-like motion can lead to an overall pusher flow field. Indeed, previously it had been shown that}
the triangular microswimmer can adopt both a puller or pusher far field, depending on  the specific swimmer shape and stroke cycle \cite{Ziegler2019}.
\az{However, so far the detailed reason and a simple explanation for observing a pusher flow field in our case are not yet clear.}





\begin{figure}
\centering
  \includegraphics[width=0.5\textwidth]{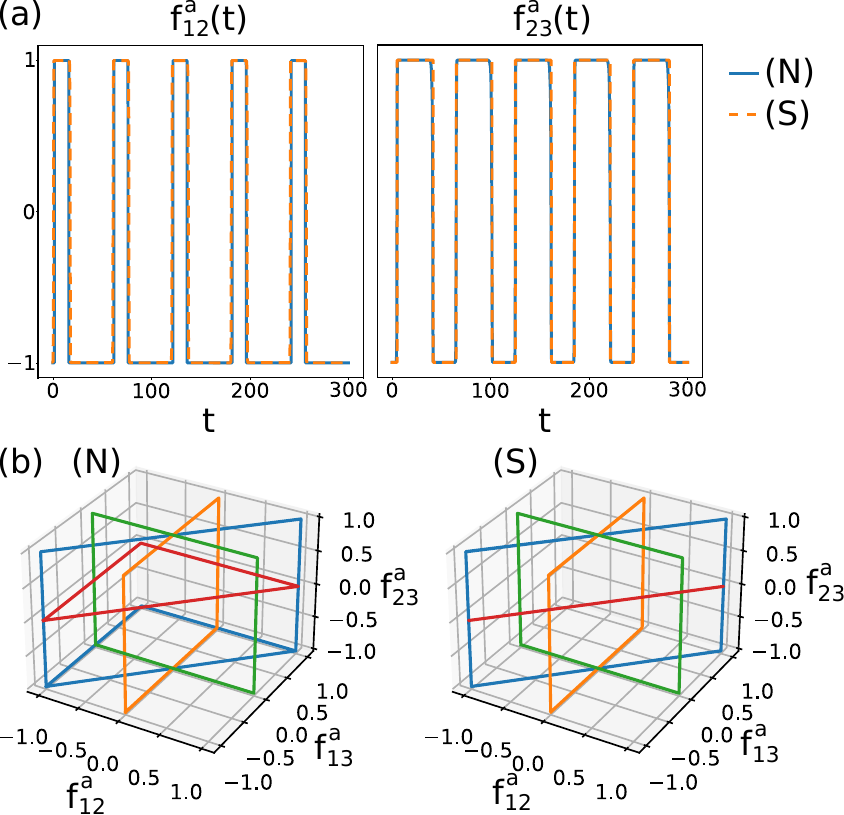} 
\caption{(a) Periodic force pattern of the arm forces $\az{f_{12}^a}(t)$ and $\az{f_{23}^a}(t)$ normalized by the maximum force $f_\text{max}$ shown for the non-symmetric (N) and symmetric swimmer for $R_b=6R_f$. (b) Motion of the arm forces $\az{f_{12}^a}(t)$, $\az{f_{13}^a}(t)$ and $\az{f_{23}^a}(t)$ in the force phase space for both (N) and (S) swimmers. \az{Blue: Actual trajectory: Red, orange, green: Projections to the 2D subspaces.}}
\label{fig:4}       
\end{figure}

\begin{figure}
\centering
  \includegraphics[width=0.82\textwidth]{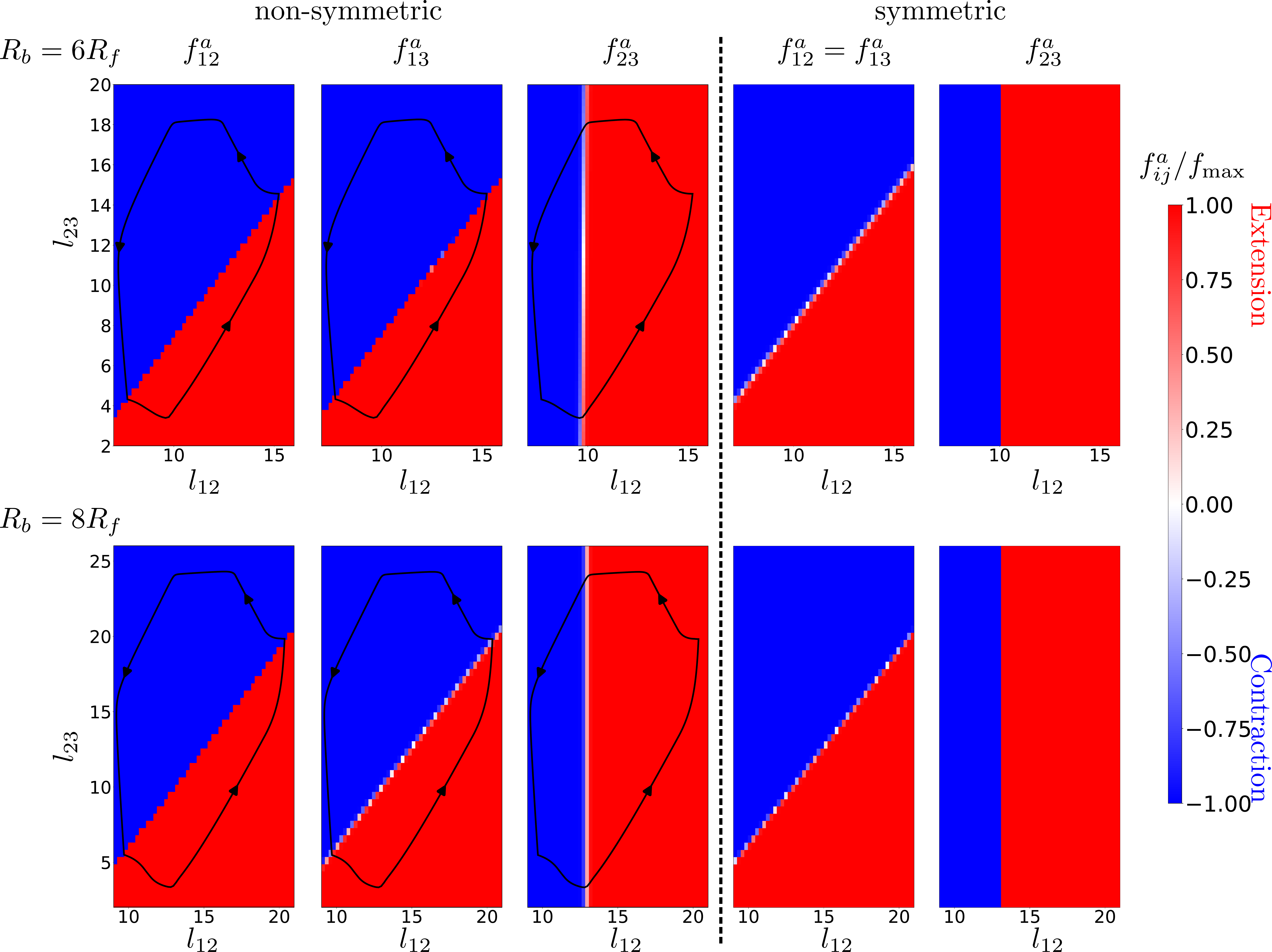} 
\caption{Quasi-optimum policies $\az{f_{ij}^a}(\{ l_{ij} \})$ obtained from Reinforcement Learning both for non-symmetric and symmetric microswimmers for two different values for the body size $R_b$. Here red colors indicate arm extensions and blue colors arm contractions. The black dashed lines indicate the dynamics of the arm lengths in $(l_{12},l_{23})$ phase space exemplarily for the non-symmetric swimmers.}
\label{fig:policy}       
\end{figure}

\subsection{Quasi-optimal policies}
\label{sec:policy}
\az{In Sec.~\ref{sec:gaits} we pointed out that the determined quasi-optimal swimming strokes are similar for the symmetric and non-symmetric cases. These}
swimming strokes 
are a result of the outcome of the arm forces $\az{f_{12}^a}$, $\az{f_{13}^a}$ and $\az{f_{23}^a}$, 
which in our setup depend only on the instantaneous shapes of the swimmer\az{s}, characterized by the\az{ir} arm lengths $l_{12}$, $l_{13}$ and $l_{23}$.
Before we discuss these policies  obtained by ANNs via the NEAT genetic algorithm (see Sec.~\ref{Sec:RL}),
we first note that the forces $\az{f_{ij}^a(t)}$ for all microswimmers are periodic with the respective stroke frequencies $f$ discussed in Sec.~\ref{sec:perform}\az{, where the forces are}
periodically switching between the extreme values $-f_\text{max}$ and $+f_\text{max}$.
As can be seen in Fig.~\ref{fig:4}(a) for $R_b=6R_f$,
\az{the quasi-optimum force patterns $f_{ij}^a(t)$  are indeed very similar for the symmetric (S) and non-symmetric (N) cases. The}
arm force\az{s} $\az{f_{23}^a}$ \az{are} phase-shifted compared to two body-flagella forces 
\az{which are in perfect synchrony for the symmetric swimmer by construction, and quasi-synchronized for the non-symmetric swimmer where }
$\az{f_{12}^a} \approx \az{f_{13}^a}$.
The phase shift of the forces leads to a periodic motion in the arm force phase space, as shown in Fig.~\ref{fig:4}(b), but where only for the synchronized swimmers, by construction, the forces $\az{f_{12}^a}$ and $\az{f_{13}^a}$ are in perfect synchrony.

The obtained quasi-optimum policies are given for the non-symmetric (N) microswimmers by three 
\az{and for the symmetric (S) microswimmer by two} nonlinear functions 
of the arm forces which depend only on the arm lengths, which can be written in shorthand notation as
$\az{f_{ij}^a}(\{ l_{ij} \})$. 
Note that these policies can be obtained by an infinite set of neural networks with different topology and weights.
In our procedure using the NEAT genetic algorithm relatively simple neural networks have been found, as exemplarily shown in Fig.~\ref{fig:B}(e).
To understand the obtained policies we show in Fig.~\ref{fig:policy} the functions $\az{f_{ij}^a}$
exemplarily for $R_b=6R_f$ and $R_b=8R_f$ both for the non-symmetric (N) and for the symmetric (S) case.
The different values of $\az{f_{ij}^a}$ are shown depending only on $l_{12}$ and on $l_{23}$ but not on $l_{13}$, because we know that the obtained solutions are almost symmetric, where $l_{13} \approx l_{12}$.

We can see that the \az{quasi-optimum} policies are rather simple,
and are very similar for non-symmetric and symmetric swimmers.
It can be seen in Fig.~\ref{fig:policy} that, as expected \az{from the results shown in Fig.~\ref{fig:4}}, the forces depending on the arm lengths are always approximately   $\approx +f_\text{max}$ (\az{extending arm force,} red) or $\approx -f_\text{max}$ (\az{contracting arm force,} blue).
The fact that the borders between regions with $+f_\text{max}$ and regions with $-f_\text{max}$ are divided by approximately straight lines allows us to write down the approximate policies \az{both for symmetric (S) and non-symmetric (N) swimmers} by
\begin{align}
    \az{f_{12}^a} \approx \az{f_{13}^a} & \approx f_\text{max}\, \text{sign}(l_{23}-c_a l_{12}-l_a) \\
    \az{f_{23}^a} & \approx f_\text{max}\, \text{sign}(l_{12}-l_b)   
\end{align}
where $l_a$, $l_b$ and $c_a$ are constants depending on $R_b$ but which are approximately the same for (N) and (S) microswimmers.
Therefore the policies can be interpreted in the following way: Always when the arm length $l_{12}$ (and hence $l_{13}$) is larger than the critical arm length $l_b$,
the flagellum-flagellum force $\az{f_{23}^a}$ is positive, which results in an extension of the flagellum-flagellum distance $l_{23}$.
In contrast, for $l_{12} < l_b$ we get $\az{f_{23}^a}<0$ leading to a contraction, $\mathrm{d} l_{23} / \mathrm{d}t <0$.
So, the body-flagellum distance governs the flagellum-flagellum contraction or extension.
In contrast, the forces between the flagella and the body, $\az{f_{12}^a}$ and $\az{f_{13}^a}$, depend on both the flagellum-body distance $l_{12}$ and on the flagellum-flagellum distance $l_{23}$: For combinations of $l_{12}$ and  $l_{23}$ located in the respective red regions in the subplots of Fig.~\ref{fig:policy} the body-flagella distances $l_{12}$ and $l_{13}$ are extended, while in the respective red regions they are contracted.
We confirm this behavior by plotting the observed dynamics \az{for the non-symmetric (N) case} in $(l_{12},l_{23})$ phase space, as indicated by the black closed curves in the subplots of Fig.~\ref{fig:policy}, which show exactly the describe contraction and extension behavior in the respective red and blue regions.
\az{The black curves look very similar for the symmetric case (S) (not shown in Fig.~\ref{fig:policy}).}







\subsection{Comparison to circular flagella-bead motion}
\begin{figure}
\centering
  \includegraphics[width=0.8\textwidth]{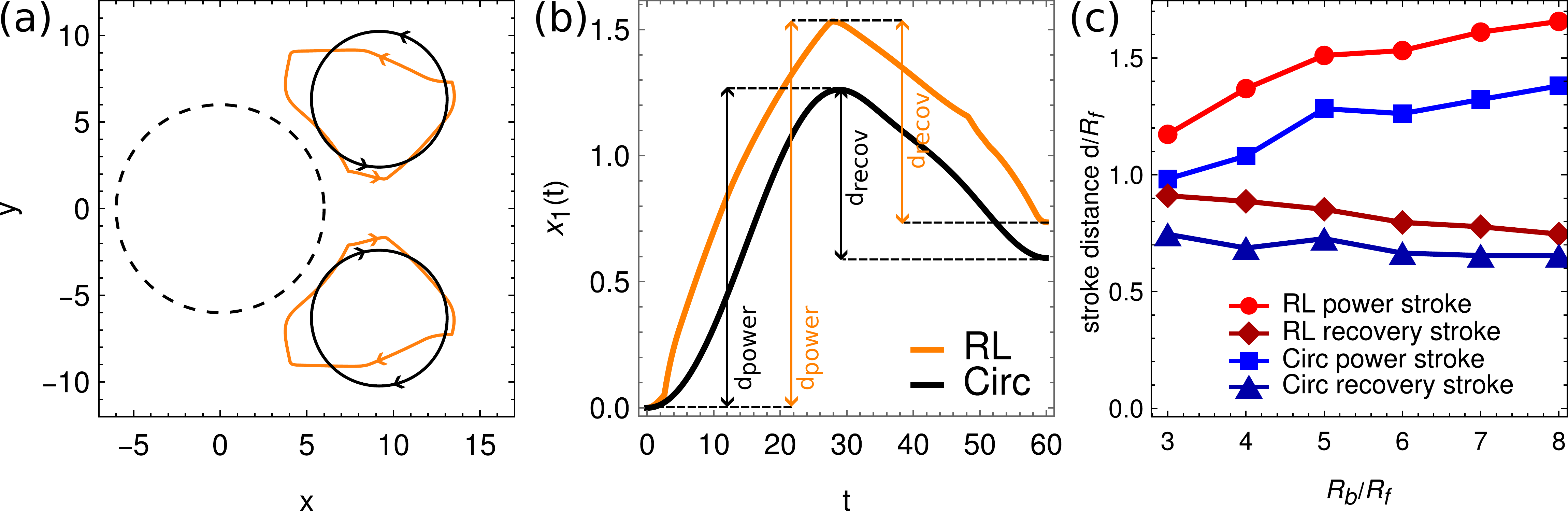} 
\caption{(a) Periodic motion of the flagella beads obtained from the RL optimization (orange curves) compared to circular flagella bead trajectories (black curves) with same enclosed area, same minimum flagella-body distance and same mean flagellum-flagellum distance, shown here for a body (black dashed curve) of radius $R_b=6R_f$. \az{The arrows indicate the direction of flagella bead motion.}
(b) Comparison of the traveled forward distance during the power stroke $d_\text{power}$ and backward distance during the recovery stroke $d_\text{recov}$ for the RL-optimized and the circular flagella bead kinematics. 
(c) Comparison of $d_\text{power}$ and $d_\text{recov}$   for different body sizes $R_b/R_f$.
}
\label{fig:cmp}       
\end{figure}

Our work uses a large body bead and two smaller flagella beads which represents biflagellates such as \textit{Chlamydomonas}, similar
as in previous works by Friedrich and coworkers \cite{Friedrich2012,Polotzek2013} and Bennett and Golestanian \cite{Bennett2013,Bennett2013b}.
In these previous works the motion of the flagella beads have been prescribed to follow circular trajectories with respect to the body position.
In contrast, in our work   
the RL optimization finds flagella bead trajectories which are not circular, as shown in Fig.~\ref{fig:cmp}(a) exemplarily for $R_b=6R_f$ (orange curves).
We determined the enclosed areas $A$ of these periodic bead patterns which depend on $R_b$ and determine corresponding effective radii $R_c = \sqrt{A/ \pi}$ for different $R_b$. 
We then compare our results from the RL optimization with simple pre-defined policies for the flagella bead motion, similar as in previous works \cite{Friedrich2012,Polotzek2013,Bennett2013,Bennett2013b}, where the flagella beads perform circular trajectories of radius $R_c$ (black circles in Fig.~\ref{fig:cmp}(a)).
The center of the circles are placed such that the mean-flagellum-flagellum distance and the minimum flagella-body distances are the same as in the RL optimization procedure. 
Furthermore, we use the same stroke frequencies as obtained from the RL optimization.
For a fixed stroke pattern it does not matter how fast the flagella beads move instantaneously along their trajectory, as long as their stroke frequency is the same,
and only the kinematics determines the fitness of the swimmer.
Hence, for  simplicity we do not prescribe the forces but the velocities of the circling flagella beads, by inverting Eq.~(\ref{eq:v}) and again implementing hydrodynamic interactions in the \az{Rotne-Prager} approximation using Eq.~(\ref{eq:M}) employing synchronously beating counter-rotating flagella beads.
The bead velocities are thus prescribed as
\begin{align}
    \mathbf{v}_1(t) &= (v_{1x}(t),0) \nonumber \\
    \mathbf{v}_2(t) &= (v_{1x}(t) + R_c \cos\omega t ,R_c \omega\sin \omega t ) \label{eq:vcirc}\\
    \mathbf{v}_3(t) &= (v_{1x}(t) + R_c \cos\omega t ,-R_c \omega\sin \omega t ) \nonumber
    \end{align}
with $\omega = 2\pi f$ using the frequencies $f$ obtained from the RL optimization (Fig.~\ref{fig:2}(a)).
The unknown body velocity $v_{1x}(t)$
is then obtained by enforcing the force-free condition \az{\textit{via}} inversion of Eq.~(\ref{eq:v}).

The black dots  in Fig.~\ref{fig:2}(b) show the determined stroke-averaged velocities obtained from the prescribed policy, $v_c = \langle v_{1x} \rangle$. 
We can see that the fitness obtained by RL outperforms the fitness with the circular flagella bead trajectories.
To further understand and quantify the differences, we measure the distances $d_\text{recov}$ in the recovery stroke and the distances $d_\text{power}$ in the power stroke, as the distances the body moves backwards and forwards in a single stroke, respectively, as shown in Fig.~\ref{fig:cmp}(b) for  $R_b=6R_f$ both for the RL-optimized and the circular flagella-bead trajectories.  
Fig.~\ref{fig:cmp}(c) shows now for different $R_b$ that the body bead in RL-optimized trajectories move backward stronger during the recovery period compared to the trajectories from the pre-defined circular system;
nevertheless the RL-optimized strategy leads to a larger net displacement because the distance traveled during the power stroke outperforms those from the pre-defined bead motion (Eq.~(\ref{eq:vcirc})). 
This is a consequence of the flagella bead trajectories as can be seen in Fig.~\ref{fig:cmp}(a):
For the RL-optimized trajectories the flagella beads stay close to the cell body for a longer time during the recovery stroke where the drag is minimized, leading to a longer recovery stroke, and larger backward motion.
On the other hand, the flagella beads have now a larger power stroke amplitude available compared to the pre-defined system, where the drag force is large, leading to larger power stroke displacement outperforming the pre-defined system
and to an overall faster swimming stroke.


\section{Conclusions}
\label{Sec:Con}
Here we applied Reinforcement Learning by using genetic algorithms to evolve topologies and weights of small artificial neural networks to predict forces $f_{ij}$ between the  beads of a biflagellate microswimmer model depending on the instantaneous bead distances $l_{ij}$ to optimize the swimming speed $v$.
In contrast to previous work  \cite{Friedrich2012,Polotzek2013,Bennett2013,Bennett2013b} the periodic motion of the flagella beads with respect to the body bead and the frequency of this periodic motion  are not predefined, but are emerging properties from the obtained policies $\az{f_{ij}^a}(\{ l_{ij} \})$.

Interestingly, we identified for this intrinsically non-symmetric (N) system for all considered ratios of body vs.~flagella bead radii $R_b/R_f$ that the swimmer  always adopts policies where the flagella beads move in synchrony, which coincide well with optimized policies obtained from our symmetric (S) system.
The fact that the policies lead to synchronized motion in this low-dimensional problem allowed us to visualize and interpret them, and to approximate them by simple analytic expressions: We found that solely the instantaneous length between body and flagella beads determines if the flagella beads stretch or contract,
while a linear relation between body-flagella and flagellum-flagellum distances determines
the instantaneous stretching/contraction of the flagella-body distances.

The obtained policies result in flagella beating patterns which deviate from previously used circular bead trajectories \cite{Friedrich2012,Polotzek2013,Bennett2013,Bennett2013b}.
While our optimized strokes enhance the backward motion of the microswimmers during the recovery stroke,
the forward motion during the power stroke is even more enhanced, leading to an overall larger net displacement within a stroke cycle compared to the pre-defined circular bead trajectories.

The instantaneous obtained flow fields are very similar as those observed experimentally during the power and recovery stroke, leading to oscillating contractile and extensile force dipole fields.
Noteworthy, our work demonstrates that breast-stroke-like motion as it occurs for the used microswimmer model can lead to a stroke-averaged  pusher type flow field, in contrast to the experimentally observed puller-type flow.
This shows that details of the stroke kinematics matters, as previously shown for simple triangular microswimmer models \cite{Ziegler2019}.

Finally we note that  we have used a purely hydrodynamic description and that noise is not included in our model.
In the future it would be interesting to investigate the effect of noise on the motion of the beads during the RL process, which can potentially change the optimized policies to achieve maximized swimming speed,
as well as on the stability of the synchronized bead motion.

%

\backmatter

\bmhead{Acknowledgements}
We thank Maximilan H\"ubl and Benedikt Hartl for useful discussions.
The computational results presented have been achieved using the Vienna Scientific Cluster (VSC).

\bmhead{Data Availability Statement}
The results and data created for this manuscript is available from the corresponding author on reasonable request.

\bmhead{Author contribution statement}
A.Z.\ conceived the study.
S.B.\ and A.Z.\ performed the simulations, analyzed the results, and wrote the paper.

\begin{appendices}

\section{Artificial Neural Networks (ANN)}
\label{A:ANN}
Artificial Neural Networks are used to approximate functions from the space of inputs to the space of outputs for a given machine learning task. The basic building block of ANNs are artificial neurons which are referred to as units. Every layer of the network consists of one or more units. Each unit takes as input all the outputs of the previous layer and creates a linear weighted combination of these \cite{MontesinosLopez2022}. A bias term is added and the resulting sum is passed through an activation function, which in our case is the hyperbolic tangent function $tanh$ for all neurons. The result of the activation function is the output of the corresponding neuron and fed into the next layer or used as output of the total ANN. In machine learning the network is optimized with respect to the weights and biases. 

\section{The NEAT Genetic Algorithm}
\label{A:NEAT}
In the NEAT neuroevolution algorithm \cite{Stanley2002a} the topology of an ANN is interpreted as a genome of a population 
where ANNs evolve by a cross-over between two genomes where a new ANN is created \az{using the concepts of}
\textit{Genetic Encoding}, \textit{Historical Markings} and \textit{Speciation} \az{as briefly} discussed \az{in the} following. For \az{details} we refer to Ref.~\cite{Stanley2002a}.

NEAT specifies \textit{genomes} which represent the network connectivity by including a list of connection genes, which refers to two connected node genes. Node genes contain a list of inputs, hidden nodes and outputs. Every connection gene describes the in- and out-node, the weight of the connection, whether the gene is expressed with the help of an enable bit and an innovation number which is used to find corresponding genes. Weights mutate with the help of stochastic perturbations and topological mutations
occur either by adding/removing connections or adding/removing nodes.
In order for a system to know which genes are aligned for \textit{crossover}, it needs to keep track of the historical origin of every gene in the system. Whenever a new structural mutation takes place, a global innovation number is assigned to that gene. When crossover takes place only genes in the two genomes with the same \textit{innovation number} are aligned for crossover.
\textit{Speciation} is a method that restricts the competition of genomes
to their own niches rather than the total population. This allows genomes to first optimize over a few generations.
It turns out that genes which are not aligned can be used to quantify a compatibility
distance, which quantifies how ”distant” two genomes are from each other \az{in order} to define different \textit{species}.
\az{Finally we note that, as} compared to other neuroevolution \az{algorithms,} NEAT starts out with a population of networks with minimal complexity and adds nodes and connections
during the evolution, \az{keeping} the search space  low \az{and thus leading} to better performance.


\end{appendices}


\end{document}